\documentstyle[preprint,revtex]{aps}
\begin{document}
\draft
\preprint{
\begin{flushright}
DPNU-92-47\\
Feb. 26, 1993
\end{flushright}
}
\vfill
\begin{title}
Non-oblique Correction in Extended Technicolor Theory
\end{title}
\vfill
\author{Noriaki Kitazawa}
\begin{instit}
Department of Physics, Nagoya University\\
Nagoya, 464-01, Japan
\end{instit}
\vfill
\begin{abstract}
We study radiative corrections on $Zb{\bar b}$ vertex
 generated by the ETC gauge bosons, ``diagonal'' as well as sideways.
Although the oblique corrections due to the ETC bosons are small
 in comparison with the oblique correction
 due to the technicolor dynamics, the non-oblique corrections
 result in substantial shift of contour plot
 in the $S$-$T$ plane.
We show that such a shift due to the non-oblique corrections
 is actually important for discussing $S$ and $T$ values
 in the technicolor models.
\end{abstract}

\newpage

The precision measurement of the electroweak observables
 gives strong constraint on the new physics beyond the standard model
 \cite{precision}.
The contribution of the new physics to the radiative correction
 can be parametrized by three parameters $S$, $T$, and $U$
 \cite{Peskin-Takeuchi}.
The precision measurement
 restricts the values of these three parameters.
It must be noted that
 this parametrization is based on the assumption
 that the non-oblique correction
 is small compared with the oblique correction \cite{Kennedy-Lynn}.

Technicolor theory \cite{technicolor},
 at least a simple QCD scale up, is strongly constrained
 by the experiments,
 because it predicts a too large $S$ \cite{Peskin-Takeuchi-recent}.
Walking technicolor \cite{walking} may predict rather small $S$,
 but it does not seem to be small enough \cite{walking-S}.
But the consideration so far is not complete,
 because the non-oblique correction
 which cannot be parametrized by $S$, $T$, and $U$ can be large.
In fact, the physics of the quark and lepton mass generation,
 particularly the large top quark mass,
 can yield large non-oblique correction \cite{Chivukula},
 which could affect the contour plot of the $S$-$T$ plane.

In the following,
 we explicitly calculate the non-oblique correction
 in a toy model of the extended technicolor theory \cite{ETC}.
We show that the non-oblique correction is indeed important
 for extracting the values of $S$ and $T$ from the experiments.

To generate the masses of the quarks and leptons,
 the technicolor gauge group is extended to the larger gauge group
 (extended technicolor (ETC) gauge group)
 which is assumed to hierarchically break down
 to the technicolor gauge group.
In the process of this breaking,
 many ETC gauge bosons become massive.
Some of them called sideways
 cause the transition of the ordinary fermions to the technifermions,
 some of them called horizontal
 connect the ordinary fermions themselves,
 and the others called ``diagonal'' diagonally interact
 with both the ordinary fermions and technifermions.
The sideways bosons
 must exist in the realistic model
 to generate the quark and lepton masses,
 while the existence of ``diagonal'' bosons is model-dependent.
The lightest bosons are the sideways and ``diagonal'' gauge bosons
 associated with the top quark.
They make the largest contributions to the radiative corrections.
In this paper, we only consider these bosons for simplicity.

Let us now consider four fundamental representations
 of the ETC gauge group $SU(N_{TC}+1)$
 containing the top and the bottom quark
 and the technifermions $U$ and $D$.
\begin{equation}
 \left(\begin{array}{cc}
  \quad
  \left(\begin{array}{c} U^1 \\ \vdots \\ U^{N_{TC}} \\ t
        \end{array}\right)_L
  \quad
  \left(\begin{array}{c} D^1 \\ \vdots \\ D^{N_{TC}} \\ b
        \end{array}\right)_L
  \quad
 \end{array}\right),
 \qquad
 \left(\begin{array}{c} U^1 \\ \vdots \\ U^{N_{TC}} \\ t
       \end{array}\right)_R,
 \qquad
 \left(\begin{array}{c} D^1 \\ \vdots \\ D^{N_{TC}} \\ b
       \end{array}\right)_R.
\end{equation}
Two left-handed representations form an $SU(2)_L$ doublet
 and two right-handed representations are $SU(2)_L$ singlets.
(We assume that the ETC gauge group commute
 with the weak interaction gauge group.)
After the breaking of ETC gauge group
 down to the technicolor gauge group $SU(N_{TC})$,
 the massive sideways and ``diagonal'' gauge bosons are generated.

In this model the mass of top quark
 is equal to the mass of bottom quark,
 because of the common mass and coupling of the sideways
 and $\langle \bar{U} U \rangle = \langle \bar{D} D \rangle$.
In the realistic model, however,
 the right-handed top quark and the right-handed bottom quark
 must be contained in the different representations
 of the ETC gauge group
 to realize the different sideways couplings.
Therefore, the ETC gauge theory must be a chiral gauge theory.
However,
 realistic representations of ETC gauge group are not known yet.
Instead of considering explicit ETC model,
 we here simply assume different ETC couplings
 for top and bottom right-handed fundamental representations,
 while keeping the technicolor interaction vector-like.
We also assume that
 $\langle \bar{U} U \rangle = \langle \bar{D} D \rangle$.
This is a good toy model for the isospin breaking,
 although the ETC gauge symmetry is destroyed.
More explicitly,
 we set the sideways coupling $\xi_t g_t$ for the left-handed quarks,
 $g_t / \xi_t$ for the right-handed top quark,
 and $g_t / \xi_b$ for the right-handed bottom quark,
 where $g_t$ is given by
\begin{equation}
 m_t \simeq {{g_t^2} \over {M_X^2}} 4 \pi F_\pi^3.
\end{equation}
The scale $M_X$ is the mass of sideways
 and the relation $\langle \bar{U} U \rangle \simeq 4 \pi F_\pi^3$
 (naive dimensional analysis) is used,
 where $F_\pi$ is the decay constant of the Nambu-Goldstone bosons
 of the technicolor chiral symmetry breaking.
Large top quark mass
 indicates large value of $g_t$ or small value of $M_X$.
We assume that the sideways effect can be treated perturbatively,
 namely, ${{(\xi_t g_t)^2} \over {4\pi}} \alt 1$
 and ${{(g_t / \xi_t)^2} \over {4\pi}} \alt 1$.
This relations restrict the value of $\xi_t$.
For the realistic bottom quark mass,
 $\xi_b$ is restricted by
 ${1 \over {\xi_b}} \leq {1 \over {\xi_t}}{{m_b} \over {m_t}}$.

The ``diagonal'' couplings
 are fixed through the relation to the sideways couplings.
For technifermion, we obtain the ``diagonal'' coupling
 by multiplying the factor
 $-{1 \over {N_{TC}}} \sqrt{{N_{TC}} \over {N_{TC} + 1}}$
 to their sideways couplings.
For quark, we obtain it by multiplying the factor
 $\sqrt{{N_{TC}} \over {N_{TC} + 1}}$
 to their sideways couplings.
These factors
 come from the normalization and traceless property
 of the diagonal generator.
The ``diagonal'' interaction is also chiral.

The sideways bosons
 yield potentially a large non-oblique correction
 to the $Z b \bar{b}$ vertex,
 which has been estimated by Chivukula et al. \cite{Chivukula}.
By using the approach of the effective Lagrangian,
 the correction to the left-handed and the right-handed couplings
 of the bottom quark are derived as \cite{Chivukula}
\begin{eqnarray}
 \delta g_L^b &=& {{\xi_t^2} \over 4}
                  {{m_t} \over {4 \pi F_\pi}} {e \over {cs}},
\nonumber\\
 \delta g_R^b &=&-{1 \over {4\xi_b^2}}
                  {{m_t} \over {4 \pi F_\pi}} {e \over {cs}},
\label{sideways}
\end{eqnarray}
 where $c$ and $s$
 are the cosine and sine of the Weinberg angle, respectively.
The suppression factor $m_t / 4\pi F_\pi$
 is not small for large $m_t$.
The diagram corresponding to this correction
 is shown in FIG.~\ref{corrections}.

The ``diagonal'' boson ``$X$'' also yields the non-oblique correction
 through the mixing with $Z$ boson \cite{Holdom}.
The mixing is parametrized by three parameters $x$, $y$, and $w$ as
\begin{eqnarray}
 {\cal L}_{AZX}
  = & - & {1 \over 4}
     \left(
       \begin{array}{ccc}
         X_{\mu\nu} & Z_{\mu\nu} & A_{\mu\nu}
       \end{array}
     \right)
     \left(
       \begin{array}{ccc}
         1 \quad    & y \quad    & w          \\
         y \quad    & 1 \quad    & 0          \\
         w \quad    & 0 \quad    & 1
       \end{array}
     \right)
     \left(
       \begin{array}{c}
         X^{\mu\nu} \\
         Z^{\mu\nu} \\
         A^{\mu\nu}
       \end{array}
     \right)
\nonumber\\
   & + & {1 \over 2}
     \left(
       \begin{array}{ccc}
         X_\mu      & Z_\mu      & A_\mu
       \end{array}
     \right)
     \left(
       \begin{array}{ccc}
         M_X^2      & x M_Z^2    & 0          \\
         x M_Z^2    & M_Z^2      & 0          \\
         0          & 0          & 0
       \end{array}
     \right)
     \left(
       \begin{array}{c}
         X^\mu \\
         Z^\mu \\
         A^\mu
       \end{array}
     \right),
\end{eqnarray}
 where we set the ``diagonal'' mass equal to the sideways mass.
Within the leading order of $x$, $y$,
 and $w$ in the four-fermion amplitude,
 the non-oblique correction to the $Zb\bar{b}$ vertex is obtained as
\begin{eqnarray}
 \delta g_L^b &=& \xi_t g_t
                  \sqrt{{N_{TC}} \over {N_{TC}+1}}
                  {{M_Z^2} \over {M_X^2-M_Z^2}} (y-x),
\nonumber\\
 \delta g_R^b &=& {{g_t} \over {\xi_b}}
                  \sqrt{{N_{TC}} \over {N_{TC}+1}}
                  {{M_Z^2} \over {M_X^2-M_Z^2}} (y-x).
\label{diagonal}
\end{eqnarray}

We get $x$, $y$, and $w$
 by calculating one-loop diagrams of FIG.~\ref{loops}
 with constant fermion mass.
The results are
\begin{equation}
 x = N_C \sqrt{{N_{TC}} \over {N_{TC} + 1}}
     {{g_t e} \over {(4\pi)^2}} {1 \over {cs}}
     \left[
           (\xi_t - {1 \over {\xi_t}} ) {{m_U^2-m_t^2} \over {M_Z^2}}
          -(\xi_t - {1 \over {\xi_b}} ) {{m_D^2-m_b^2} \over {M_Z^2}}
     \right],
\end{equation}
\begin{eqnarray}
 y = &-& N_C \sqrt{{N_{TC}} \over {N_{TC} + 1}}
     \xi_t
     {{g_t e} \over {(4\pi)^2}} {1 \over {cs}} {1 \over 3}
     \left[
            \ln {{m_t^2} \over {m_b^2}} - \ln {{m_U^2} \over {m_D^2}}
     \right]
\nonumber\\
     &-& N_C \sqrt{{N_{TC}} \over {N_{TC} + 1}}
     {{g_t e} \over {(4\pi)^2}} {s \over c} {2 \over 3}
     \left[
           {2 \over 3} (\xi_t + {1 \over {\xi_t}} )
            \ln {{m_U^2} \over {m_t^2}}
          -{1 \over 3} (\xi_t + {1 \over {\xi_b}} )
            \ln {{m_D^2} \over {m_b^2}}
     \right],
\end{eqnarray}
\begin{equation}
 w = N_C \sqrt{{N_{TC}} \over {N_{TC} + 1}}
     {{g_t e} \over {(4\pi)^2}} {2 \over 3}
     \left[
           {2 \over 3} (\xi_t + {1 \over {\xi_t}} )
            \ln {{m_U^2} \over {m_t^2}}
          -{1 \over 3} (\xi_t + {1 \over {\xi_b}} )
            \ln {{m_D^2} \over {m_b^2}}
     \right].
\end{equation}
Because the diagonal generator is traceless,
 the kinetic mixing parameters $y$ and $w$ are naturally finite.
The mass mixing parameter $x$ should be naturally finite
 if we use the dynamical fermion mass having momentum dependence.
To include the effect of dynamical mass,
 we set the momentum cutoff to the fermion mass
 in the individual loops, and get a finite $x$.
The diagram corresponding to the correction of eq.(\ref{diagonal})
 is shown in FIG.~\ref{corrections}.

The effect of technicolor dynamics (resonance effect)
 is not included in this one-loop calculation of $x$, $y$, and $w$.
However, we expect that
 this ambiguity might give rise to only a factor difference.
Actually,
 in the estimation of the oblique correction
 due to the technicolor dynamics,
 the $S$ parameter estimated by the QCD scale up
 is just twice the one estimated by the one-loop calculation
 \cite{Peskin-Takeuchi}.
Pseudo-Nambu-Goldstone bosons
 might also generate non-oblique correction
 which is expected to be not as large as to affect our conclusion.

The $\xi_t$ dependence of the sideways contribution
 to the left-handed coupling is quadratic and strong.
As for the ``diagonal'' contribution, it is approximately quadratic,
 but the dependence is very weak
 and the contribution is almost constant
 within the possible region of $\xi_t$.
Both contributions are positive and do not cancel each other.
The ``diagonal'' contribution
 is $30$\% of the sideways contribution when $\xi_t=1$.
We can approximately write the ``diagonal'' contribution as
\begin{equation}
 \delta g^b_L = {1 \over {(4\pi)^{4/3}}}
                {{N_C N_{TC}} \over {N_{TC}+1}}
                {{m_t} \over {4\pi F_\pi}} {e \over {cs}}
\end{equation}
 which should be compared with eq.(\ref{sideways}) of $\xi_t=1$.

Both sideways and ``diagonal'' bosons
 also contribute to the oblique correction, $S$, $T$ and $U$.
However, since the contribution is the two-loop effect,
 it is small in comparison with the contribution
 due to the technicolor dynamics.
For example,
 the ``diagonal'' contribution to the $S$ parameter is $S=0.022$
 for $M_X=1\ $TeV, $F_\pi=125\ $GeV, $N_{TC}=4$, $\xi_t=1$,
 $m_t=150\ $GeV, and $M_H=1\ $TeV,
 which is small in comparison with the contribution $S \agt 0.4$
 due to the technicolor dynamics.

Now, we observe that
 the non-oblique corrections eq.(\ref{sideways}) and (\ref{diagonal})
 change the values of $S$, $T$, and $U$
 which are extracted from the experiment.
For example, the total $Z$ width can be written as
\begin{equation}
 \Gamma_Z = \Gamma_Z^{SM} + \delta \Gamma_Z + a S + b T,
\end{equation}
 where $a$ and $b$ are certain coefficients,
 $\Gamma_Z^{SM}$ is the one-loop prediction of the standard model,
 and
\begin{equation}
 \delta \Gamma_Z
  = {{\partial \Gamma_Z^{SM}} \over {\partial g_L^b}} \delta g_L^b
  + {{\partial \Gamma_Z^{SM}} \over {\partial g_R^b}} \delta g_R^b.
\end{equation}
Therefore,
 the allowed region in the $S$-$T$ plane ($S$-$T$-$U$ space)
 by $\Gamma_Z$ is shifted by $\delta\Gamma_Z \neq 0$,
 and the resultant elliptical (ellipsoidal) region
 which is favored by the several selected experiments
 (including $\Gamma_Z$) is moved.
The predictions for the R ratio on $Z$ pole
 and forward-backward asymmtery of $b$ quark
 are also changed by the non-oblique corrections
 eq.(\ref{sideways}) and (\ref{diagonal})
 in the same way as $\Gamma_Z$,
 which also results in the shift of resultant ellipse (ellipsoid).

The contours in the $S$-$T$ plane is shown in FIG.\ref{contour}.
We set $M_X=1\ $TeV, $F_\pi=125\ $GeV (one family model),
 $m_U=m_D=(4\pi F_\pi^3)^{1/3}$,
 $N_{TC}=4$, $\xi_t=1$, $m_t=150\ $GeV, and $M_H=1\ $TeV.
The possible region of $\xi_t$ is $0.7 \alt \xi_t \alt 1.4$.
The following experiments are considered;
 total $Z$ width, $R$ ratio on $Z$ pole,
 forward-backward asymmetry of $b$ and $\mu$,
 polarization asymmetry of $\tau$,
 deep inelastic neutrino scattering ($g_L$ and $g_R$),
 and atomic parity violation $Q_W(^{133}_{\ 53}C_S)$
 \cite{experiments}.

As we take large value of $\xi_t$,
 the center of the contour moves towards the large values
 of $S$ and $T$.
This means that effectively large oblique correction
 is partially canceled by the non-oblique correction.
Both sideways and ``diagonal'' contributions
 to the right-handed $Zb\bar{b}$ coupling are suppressed
 by the power of ${1 \over {\xi_b}}$.
Namely,
 these contributions are suppressed
 by the power of ${{m_b} \over {m_t}}$
 compared with the contributions to the left-handed coupling.
Therefore,
 the contributions to the right-handed coupling are not important.

There is a difficulty of goodness-of-fit, however.
Because the correction affects only on the experiments
 related with $Zb\bar{b}$ vertex,
 the value of $\chi^2_{min}$ ($\chi^2$ at the center of contour)
 becomes large and the goodness-of-fit becomes worse.
Namely,
 the contour based on the experiments related with $Zb\bar{b}$ vertex
 do not overlap with the contour based on the other experiments.
This means that the extended technicolor theory
 (in which the ETC gauge group and weak interaction gauge group
 commute)
 can be inconsistent with the experiments.
This difficulty may disappears in the progress on the experiment.

In conclusion, we explicitly calculated the radiative correction
 generated by the extended technicolor gauge bosons in a toy model.
The non-oblique contribution
 due to the sideways and ``diagonal'' bosons
 (which are singlets of weak interaction gauge group by assumption)
 cannot be negligible in comparison with the oblique correction
 due to the technicolor dynamics.
The correction results in the shift of contour in $S$-$T$ plane.
When we extract the values $S$ and $T$ from the experiments,
 the non-oblique correction, which has as far been neglected,
 must be considered.
Actually, as we have demonstrated in this paper,
 the effect of the non-oblique correction
 can partially cancel the large oblique correction.
Thus we cannot exclude the technicolor theory
 solely through the discussions of the oblique correction.

\acknowledgments
I am grateful to K.Yamawaki and M.Harada
 for helpful discussion and advice.

\figure{
The diagrams of the non-oblique correction to the $Zb\bar{b}$ vertex:
(a) Sideways contribution. (b) ``Diagonal'' contribution.
\label{corrections}
}

\figure{
The one-loop diagrams for calculating $X$-$W_3$ and $X$-$B$ mixing.
\label{loops}
}
\figure{
The region favored by the experiments.
The single contour indicates the favored region (68\% C.L.)
 when the sideways and ``diagonal'' correction is absent.
The contours in bold line (68\% and 90\% C.L.)
 indicate the favored region with the correction ($\xi_t=1$).
The prediction
 of the technicolor dynamics ($N_{TC}=4$) for $S$ parameter
 is shown in doted lines for one doublet ($S \simeq 0.4$)
 and one family ($S \simeq 1.6$) model.
\label{contour}
}

\end{document}